# Measurement of the overlap between quantum states with the use of coherently addressed teleportation

Andrzej Grudka* and Antoni Wójcik**

Faculty of Physics, Adam Mickiewicz University,

Umultowska 85, 61-614 Poznań, Poland

Abstract

We will show how to measure the overlap between photon polarization states with the use of linear optics and postselection only. Our scheme is based on quantum teleportation and succeeds with the probability of $1/8$.

PACS number(s): 03.67.-a

Many concepts of quantum information processing have been first implemented with the use of linear optics. Among them there are e.g. quantum teleportation [1], dense coding [2], quantum cryptography (e.g. [3, 4]). These schemes use polarization degrees of freedom of single photons for encoding qubits. Unfortunately, linear optics does not allow to perform arbitrary two-qubit operations in a deterministic way. However, it was shown theoretically [5, 6] as well as experimentally [7, 8] that it is possible to achieve this aim in a nondeterministic way. The main idea is to use quantum measurement as a source of nonlinearity. For example, one can construct CNOT gate which succeeds with the probability of $1/4$ [9]. As well known

[10] CNOT gate together with one-qubit gates are universal, i.e. they can be used to implement an arbitrary unitary operation.

Recently, some authors [11-16] have pointed out the usefulness of the quantum circuit presented in Fig. 1. This scheme consists of two Hadamard gates H, phase gate $\varphi$ and one controlled unitary operation U. It allows one to estimate $Tr(U\rho)$. In particular, for the measurement of the overlap between two quantum states $|\langle\varphi|\psi\rangle|$ one removes the phase gate and takes SWAP gate (defined as follows: $SWAP|\varphi\rangle|\psi\rangle=|\psi\rangle|\varphi\rangle$) instead of U. It is the aim of this paper to present the scheme which allows a measurement of the overlap between the polarization states of two photons with the use of linear optics. Of course, the C-SWAP gate (as any gate) can be constructed from two-qubits gates (e.g. five two-qubit gates as proposed by Smolin and DiVincenzo [17]), which can be independently performed with the probability of success bounded from the above by $1/4$. So the measurement of the overlap based on this naive approach to the construction of C-SWAP gate succeeds with the probability bounded by $4^{-5}\approx0.001$. In contrast, the scheme presented here has the probability of success equal to $1/8$.

Let us now present the C-SWAP gate in a slightly different way. This gate (Fig. 2) involves five spatial modes: one control ($C$), two input (1, 2) and two output (3, 4) modes. The polarization states of photons in the input modes are transmitted to different output modes. Obviously, there are two possible ways of such a transmission. Which one of them is taken depends on the polarization state of the photon in the control mode. Namely, if the control photon is in the state $|H\rangle$ then the states of the photons in modes 1 and 2 are transmitted to modes 3 and 4, respectively. However, if the control photon is in the state $|V\rangle$ then the state of the photon in mode 1 is transmitted to mode 4 and the state of the photon in mode 2 is transmitted to mode 3. In particular, this transmission can be performed with the use of quantum teleportation (Fig. 3). In this case, one needs two teleportation channels. Each channel consists of a pair of photons in the Bell state $|\Psi^+\rangle_{\mu,\nu'}$, where

$$|\Psi^\pm\rangle_{\mu,\nu'} = \frac{1}{\sqrt{2}}(|H\rangle_\mu|V\rangle_{\nu'} \pm |V\rangle_\mu|H\rangle_{\nu'}). \qquad (1)$$

If the control photon is in the state $|H\rangle$ ($|V\rangle$), the quantum channel should be of the form $|\Psi^+\rangle_{3,3'}|\Psi^+\rangle_{4,4'}$ ($|\Psi^+\rangle_{3,4'}|\Psi^+\rangle_{4,3'}$). In the particular case of the measurement of the overlap between two photon polarization states, the control photon is in the state $\frac{1}{\sqrt{2}}(|H\rangle_C + |V\rangle_C)$ so we perform teleportation with the use of the following state:

$$|C-SWAP\rangle = \frac{1}{\sqrt{2}}\left(|\Psi^+\rangle_{3,3'}|\Psi^+\rangle_{4,4'}|H\rangle_C + |\Psi^+\rangle_{3,4'}|\Psi^+\rangle_{4,3'}|V\rangle_C\right). \qquad (2)$$

(The above state does not allow execution of C-SWAP operation for arbitrary state of the control photon.) Let us now suppose that we perform two Bell measurements, the first one on photons in modes 1 and 3', and the second one on photons in modes 2 and 4'. Eq. 2 then tells us that the state of the photon in the mode 1 is teleported to the mode 3 or 4, depending on the state of the control photon. For example, if the control photon is found in the state $|H\rangle$ then the state of the photon in the mode 1 appears in the mode 3. One can thus think that the state of the control photon plays the role of the teleportation address. Of course, the construction of the state $|C-SWAP\rangle$ ensures that the states of photons in modes 1 and 2 are always teleported to different modes. The following equations allow us to see more clearly how the above scheme works. The states of photons in modes 1 and 2 are $|\varphi\rangle$ and $|\psi\rangle$, respectively, where

$$|\varphi\rangle = a|H\rangle + b|V\rangle \qquad (3)$$

and

$$|\psi\rangle = c|H\rangle + d|V\rangle. \tag{4}$$

Thus the state of the whole system, consisting of two input qubits (modes 1 and 2), control qubit (mode $C$) and four auxiliary qubits (modes 3, 3', 4 and 4'), is

$$|\varphi\rangle_1|\psi\rangle_2|C-SWAP\rangle = |\varphi\rangle_1|\psi\rangle_2 \frac{1}{\sqrt{2}}\left(|\Psi^+\rangle_{3,3'}|\Psi^+\rangle_{4,4'}|H\rangle_C + |\Psi^+\rangle_{3,4'}|\Psi^+\rangle_{4,3'}|V\rangle_C\right). \tag{5}$$

Because we are interested in Bell measurements we will write the above state in a more convenient way where the states of photons in modes 1 and 3' as well as photons in modes 2 and 4' are expressed in the Bell basis. The resulting state consists of 16 orthogonal terms

$$\begin{aligned}|\varphi\rangle_1|\psi\rangle_2|C-SWAP\rangle = &\frac{1}{4}\sum_{\pm}\sum_{\pm'}|\Psi^\pm\rangle_{1,3'}|\Psi^{\pm'}\rangle_{2,4'}|\chi^{\pm,\pm'}_{\Psi\Psi}\rangle_{3,4,C} + \\ &+\frac{1}{4}\sum_{\pm}\sum_{\pm'}|\Psi^\pm\rangle_{1,3'}|\Phi^{\pm'}\rangle_{2,4'}|\chi^{\pm,\pm'}_{\Psi\Phi}\rangle_{3,4,C} + \\ &+\frac{1}{4}\sum_{\pm}\sum_{\pm'}|\Phi^\pm\rangle_{1,3'}|\Psi^{\pm'}\rangle_{2,4'}|\chi^{\pm,\pm'}_{\Phi\Psi}\rangle_{3,4,C} + \\ &+\frac{1}{4}\sum_{\pm}\sum_{\pm'}|\Phi^\pm\rangle_{1,3'}|\Phi^{\pm'}\rangle_{2,4'}|\chi^{\pm,\pm'}_{\Phi\Phi}\rangle_{3,4,C} +\end{aligned} \tag{6}$$

where $|\Psi^\pm\rangle$ was defined in Eq. 1,

$$|\Phi^\pm\rangle_{\mu,\nu'} = \frac{1}{\sqrt{2}}\left(|H\rangle_\mu|H\rangle_{\nu'} \pm |V\rangle_\mu|V\rangle_{\nu'}\right) \tag{7}$$

and $|\chi\rangle$ are some normalized states of three photons in modes 3, 4 and $C$. We will not present explicit form of all states $|\chi\rangle$ because we will only use two of them, namely $|\chi^{++}_{\Psi\Psi}\rangle$ and $|\chi^{--}_{\Psi\Psi}\rangle$. Because linear optics allows a discrimination of only two of four Bell states i.e. $|\Psi^\pm\rangle$ we cannot make use of the three last sums in Eq. 6. The reason for which we also ignore two terms in the first sum is much more subtle and will become apparent later. For clarity, let us write once more the state given by Eq. 6 in the form emphasizing useful terms

$$|\varphi\rangle_1|\psi\rangle_2|C-SWAP\rangle = \frac{1}{4}|\Psi^+\rangle_{1,3'}|\Psi^+\rangle_{2,4'}\frac{1}{\sqrt{2}}\left(|\varphi^+\rangle_3|\psi^+\rangle_4|0\rangle_5 + |\psi^+\rangle_3|\varphi^+\rangle_4|1\rangle_C\right) +$$
$$+\frac{1}{4}|\Psi^-\rangle_{1,3'}|\Psi^-\rangle_{2,4'}\frac{1}{\sqrt{2}}\left(|\varphi^-\rangle_3|\psi^-\rangle_4|0\rangle_C + |\psi^-\rangle_3|\varphi^-\rangle_4|1\rangle_C\right) + \quad , \quad (8)$$
$$+ \ the \ other \ terms$$

where

$$|\varphi^+\rangle = a|H\rangle + b|V\rangle = |\varphi\rangle$$

$$|\varphi^-\rangle = a|H\rangle - b|V\rangle = Z|\varphi\rangle$$

$$|\psi^+\rangle = c|H\rangle + d|V\rangle = |\psi\rangle \quad (9)$$

$$|\psi^-\rangle = c|H\rangle - d|V\rangle = Z|\psi\rangle.$$

If one performs two Bell measurements and obtains as a result $|\Psi^+\rangle|\Psi^+\rangle$ or $|\Psi^-\rangle|\Psi^-\rangle$, then the state of the remaining photons is

$$|\chi_{\Psi\Psi}^{++}\rangle = \frac{1}{\sqrt{2}}\left(|\varphi^+\rangle_3|\psi^+\rangle_4|0\rangle_C + |\psi^+\rangle_3|\varphi^+\rangle_4|1\rangle_C\right) \quad (10)$$

or

$$|\chi_{\Psi\Psi}^{--}\rangle = \frac{1}{\sqrt{2}}\left(|\varphi^-\rangle_3|\psi^-\rangle_4|0\rangle_C + |\psi^-\rangle_3|\varphi^-\rangle_4|1\rangle_C\right), \quad (11)$$

respectively.

One can see that the state $|\chi_{\Psi\Psi}^{++}\rangle$ is what we should expect as a result of C-SWAP operation performed on three photons in the state $\frac{1}{\sqrt{2}}\left(|H\rangle_C + |V\rangle_C\right)|\varphi\rangle|\psi\rangle$, i.e.

$$|\chi_{\Psi\Psi}^{++}\rangle = C - SWAP\left(\frac{1}{\sqrt{2}}\left(|H\rangle_C + |V\rangle_C\right)|\varphi\rangle|\psi\rangle\right). \quad (12)$$

Furthermore, the state $|\chi_{\Psi\Psi}^{--}\rangle$ can be transformed into the state $|\chi_{\Psi\Psi}^{++}\rangle$ by one-photon unitary operations. On the other hand, the states $|\chi_{\Psi\Psi}^{-+}\rangle$ and $|\chi_{\Psi\Psi}^{+-}\rangle$, which appear when the results of Bell measurements are $|\Psi^-\rangle|\Psi^+\rangle$ and $|\Psi^+\rangle|\Psi^-\rangle$, cannot be transformed into the state $|\chi_{\Psi\Psi}^{++}\rangle$ by the same means and are thus useless. So one is forced to accept only two results of Bell measurements, namely $|\Psi^+\rangle|\Psi^+\rangle$ and $|\Psi^-\rangle|\Psi^-\rangle$. Each of them appears with the probability of $1/16$, so the total probability of obtaining the state $|\chi_{\Psi\Psi}^{++}\rangle$ with the use of linear optics and postselection equals to $1/8$.

The second Hadamard transform (Fig. 1) is equivalent to the change of basis in which the control photon is measured, i.e. $|0\rangle \to |+\rangle$ and $|1\rangle \to |-\rangle$, where

$$|\pm\rangle = \frac{1}{\sqrt{2}}(|H\rangle \pm |V\rangle). \tag{13}$$

Eq. 10 in this new basis has the form

$$|\chi_{\Psi\Psi}^{++}\rangle = \frac{1}{2}(|\varphi^+\rangle_3|\psi^+\rangle_4 + |\psi^+\rangle_3|\varphi^+\rangle_4)|+\rangle_C + \frac{1}{2}(|\varphi^+\rangle_3|\psi^+\rangle_4 - |\psi^+\rangle_3|\varphi^+\rangle_4)|-\rangle_C. \tag{14}$$

The measurement performed on the control photon gives us the results $|+\rangle$ and $|-\rangle$ with the probabilities

$$P_+ = \frac{1}{2}(1 + |\langle\varphi|\psi\rangle|^2)$$
$$P_- = \frac{1}{2}(1 - |\langle\varphi|\psi\rangle|^2) \tag{15}$$

respectively. Thus the aim of the above scheme was achieved.

Let us also notice that the measurement performed on the state $|\chi_{\Psi\Psi}^{--}\rangle$ instead of $|\chi_{\Psi\Psi}^{++}\rangle$ gives the results $|+\rangle$ and $|-\rangle$ with the same probabilities. Thus, if one restricts the use of C-SWAP gate to the measurement of the overlap between two states, the transformation of the state $|\chi_{\Psi\Psi}^{--}\rangle$ into $|\chi_{\Psi\Psi}^{++}\rangle$ can be omitted.

In conclusion, we have presented a method allowing measurement of the overlap between photon polarization states using only linear elements. The measurement succeeds with the probability of $1/8$.

We would like to thank the Polish Committee for Scientific Research for financial support under grant no. 0 T00A 003 23.

*Email address: agie@amu.edu.pl

**Email address: antwoj@amu.edu.pl

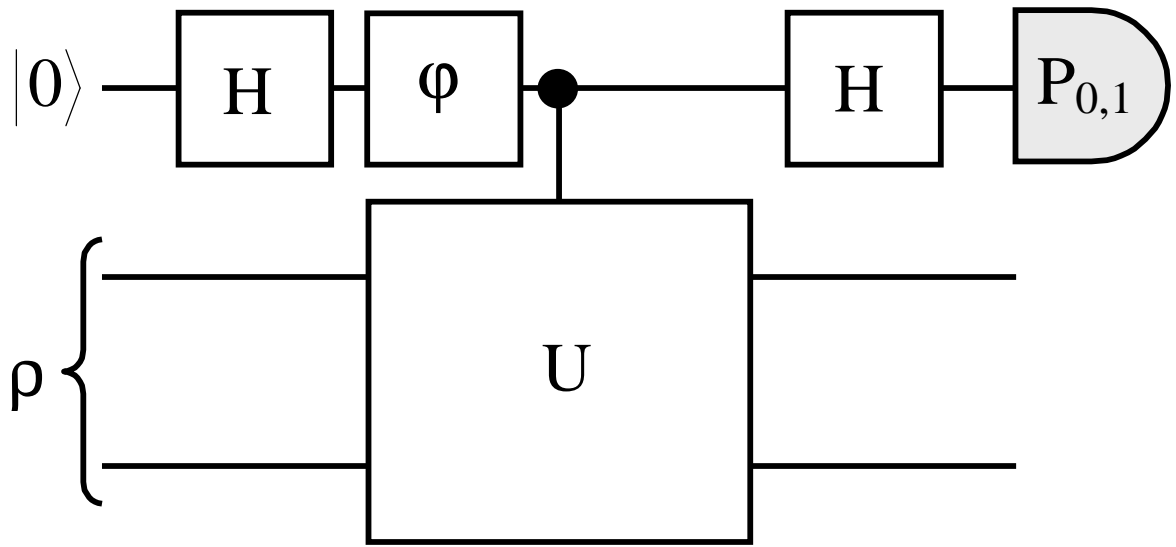

Fig. 1. Quantum circuit which allows one to estimate $Tr(U\rho)$. H is Hadamard gate, $\varphi$ is phase gate and $P_{0,1}$ is a measurement in the computational basis. $\rho$ is density matrix of two-qubit state.

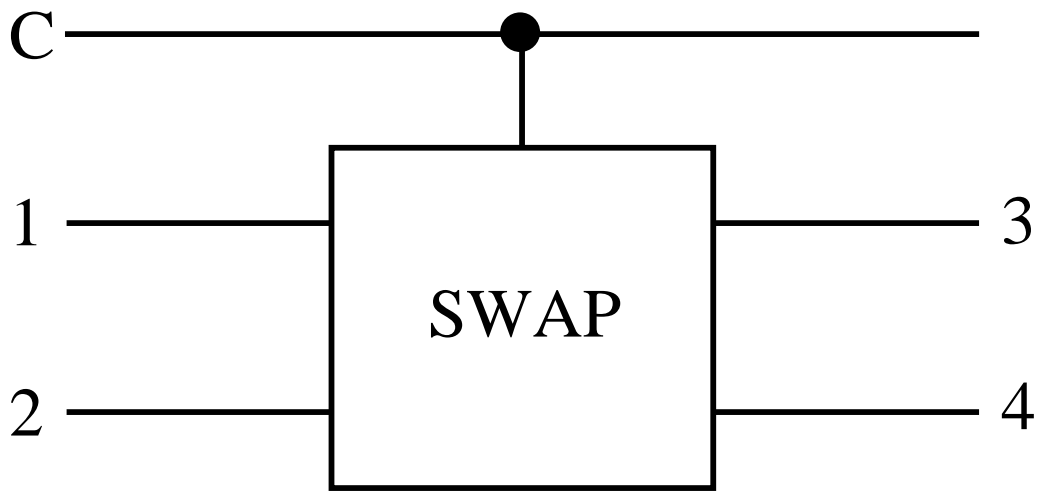

Fig. 2. Controlled SWAP gate. C is control mode, 1, 2 are input modes and 3, 4 are output modes.

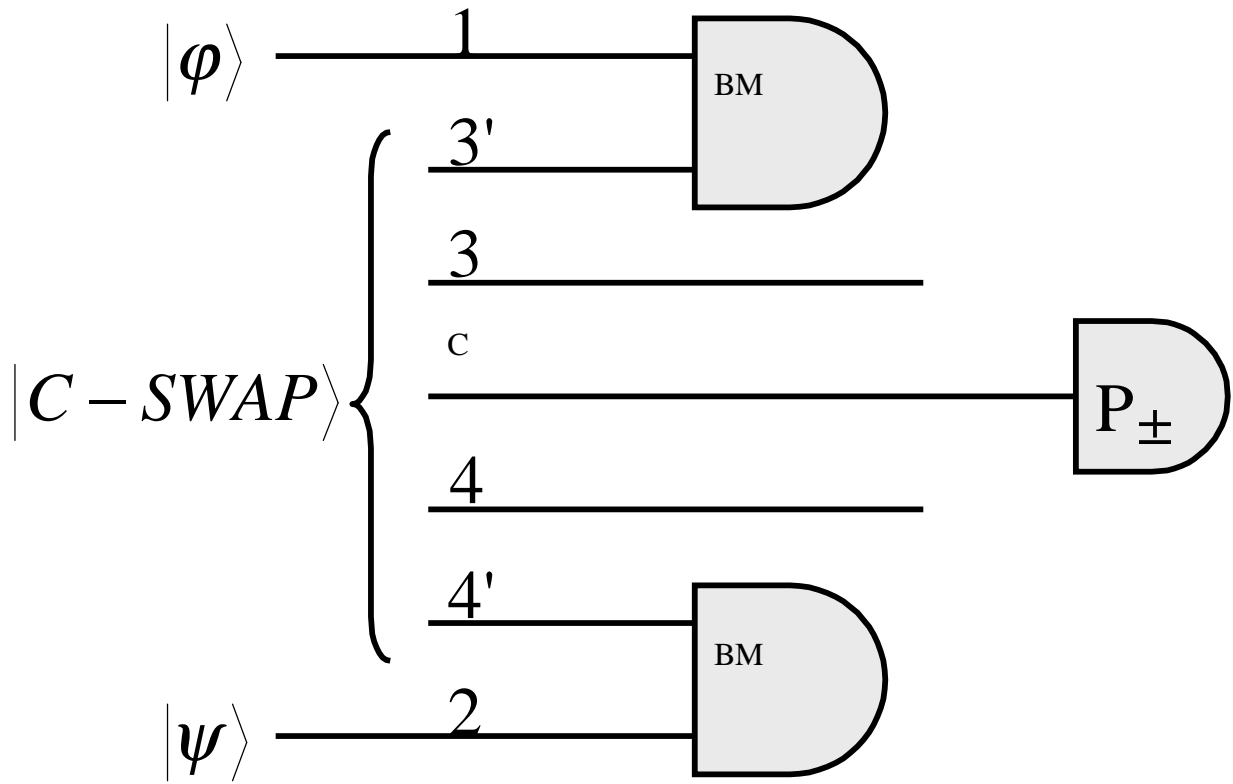

Fig. 3. Optical circuit which allows one to measure the overlap between two quantum states $|\varphi\rangle$ and $|\psi\rangle$. $|C-SWAP\rangle$ is five-photon auxiliary state. BM is Bell measurement, $P_\pm$ is measurement in $|\pm\rangle$ basis.